\documentclass[10pt, conference]{IEEEtran}

\usepackage{booktabs}   
\usepackage{subcaption} 
\usepackage{graphicx}
\usepackage{url} 
\usepackage{comment}
\usepackage{flushend}

\usepackage{ifthen}
\usepackage{color}
\usepackage{xspace}
\usepackage{cite}
\newboolean{comments}
\setboolean{comments}{true}
\newcommand\todo[1]{\ifcomments{\color{red}{\bf ToDo: #1}}\fi}

\newcommand{\acronym}{LaKe\xspace}

\begin{document}
\bstctlcite{IEEEexampleBSTcontol}

\title{\acronym: An Energy Efficient, Low Latency, Accelerated Key-Value Store} 

\author{
	\IEEEauthorblockN{ Yuta Tokusashi }
	\IEEEauthorblockA{
		Keio University\\
		tokusasi@arc.ics.keio.ac.jp
	}
	\and
	\IEEEauthorblockN{ Hiroki Matsutani }
	\IEEEauthorblockA{
		Keio University\\
		tokusasi@arc.ics.keio.ac.jp
	}
	\and
	\IEEEauthorblockN{ Noa Zilberman }
	\IEEEauthorblockA{
		University of Cambridge\\
		Noa.Zilberman@cl.cam.ac.uk
	}
}

\maketitle

\begin{abstract}
Key-value store is a popular type of cloud computing applications. The performance of key-value store applications have been shown to be very sensitive to load within the data center, and in particular to latency. As load within data center increases, it is becoming hard to maintain key-value store applications' performance, without exceeding both the processing capacity of hosts and the power budgets of racks.  
In this paper, we present \acronym: a low latency, power efficient  
key-value store design for cloud applications.
\acronym is a modular design, combining multiple cores and cache layering, both in hardware and software.
\acronym achieves full line rate throughput, while maintaining a latency of 1.1$\mu$s and better power efficiency than existing hardware based memcached designs.
Using the modularity of our design, we study trade-offs in the use of on-chip memory, SRAM and DRAM in accelerated designs and provide insights for future architectures.

\end{abstract}


\IEEEpeerreviewmaketitle

\section{Introduction\label{sec:intro}}
Online services such as e-commerce and social networks 
have nowadays become part of everyday life. These services, mostly running in the cloud\cite{DeCandia_SOSP'07}, are commonly using key-value store (KVS). 
KVS deployments in datacenters are often scaled-out in order to increase performance\cite{Rajesh_NSDI13}, which leads in turn to an increased power consumption. 
One of the limitations of KVS is that it is very sensitive to latency, in the order of tens of microseconds, end-to-end\cite{Noa_PAM'17}. Reducing latency and improving power efficiency are therefore of an utmost importance for KVS based applications.

The speed gap between networking and computation has been continuously increasing and is no longer negligible\cite{Noa_RSTA'16}.
IO-intensive applications, such as KVS, are becoming CPU-bound. To attend to this problem, recent research explored both software-based acceleration\cite{Lim_NSDI'14, Li_ISCA'15, Li_TOCS'16, Li_MICRO'16, Lim_SIGMOD'17, Liu_ASPLOS'17, Shim_JSS'17} and hardware-based 
acceleration\cite{Chalamarasetti, Nik_ATC'17, Labasani, Jae, Blott13a, Blott15, Tokusashi_HOTI'16, Tokusashi_MICRO'17, Lim, John, Fukuda, Xu_VLDB'16}. Software-based acceleration used dedicated IO libraries, (e.g., Intel DPDK), moving the application closer to the network, bypassing the kernel's network stack and processing KVS on kernel module using netfilter~\cite{Xu}.

Hardware acceleration offloads applications to a network device, e.g., a network interface card (NIC) or a switch, equipped with a dedicated FPGA/ASIC. The role of the FPGA/ASIC is run as much computation as possible in the network (or other programmable) hardware instead of on the CPU. Moving the computation to the hardware also moves power consumption from the CPU to the acceleration platform. 
Previous work~\cite{Tokusashi_HOTI'16, Tokusashi_MICRO'17} proposed multilevel NOSQL cache as a mean to improve throughput, but did not implement a design capable of full line rate. 
While both software and hardware based accelerations increase the performance, the energy efficiency of these solutions was not thoroughly studied~\cite{Tokusashi_HOTI'16, Tokusashi_MICRO'17}.

In this paper, we present \acronym: an energy efficient, low latency, accelerated KVS.
\acronym is a highly modular design, using multiple cache layers, combined with multiple processing cores, to achieve high throughput. Furthermore, it concurrently supports normal NIC operation, eliminating the need for dedicated hardware.
Unlike other specialized designs~\cite{Li_SOSP'17}, \acronym's building blocks conform with memcached and do not require a specialized application.

\acronym explores trade-offs in design and performance by leveraging multiple types of on-chip and on-board memories: on-board DRAM as a large  data store, on-board SRAM for slab allocation and on-die memory for caching and concealing latency of the external memory devices. 
\acronym  further uses a multi-processor architecture to explore scalability and latency trade-offs. 
\acronym is implemented on the NetFPGA-SUME platform~\cite{netfpga-sume}, and detailed throughput, latency and power consumption evaluations are provided, as well as a comparison to the state-of-the-art in KVS acceleration.

In this paper we make the following contributions:
\begin{itemize}
  \item We introduce \acronym: a modular architecture for KVS acceleration, using multiple processing cores, several layers of cache and hardware/software integration.
  \item We provide a detailed evaluation of \acronym, implemented on the NetFPGA-SUME platform, and show that it can reach full line rate, while providing 1.1 $\mu$s latency and $\times$5.1 better power efficiency than an existing hardware based memcached system.
  \item We explore trade-offs in using multiple layers of cache, both on-chip, on-board and on-host, and show their impact on performance and power efficiency.
\end{itemize}

The rest of this paper is organized as follows. We first describe
\acronym's architecture in Section~\ref{sec:arch}.
Section~\ref{sec:imp} describes the integration of \acronym on NetFPGA-SUME platform.
We evaluate \acronym in Section~\ref{sec:eval}.
Section~\ref{sec:rwork} explores related work
and we conclude 
in Section~\ref{sec:conc}.

\section{Architecture\label{sec:arch}}
%
%
\begin{figure}[t]
  \begin{center}
  \includegraphics[width=1\linewidth]{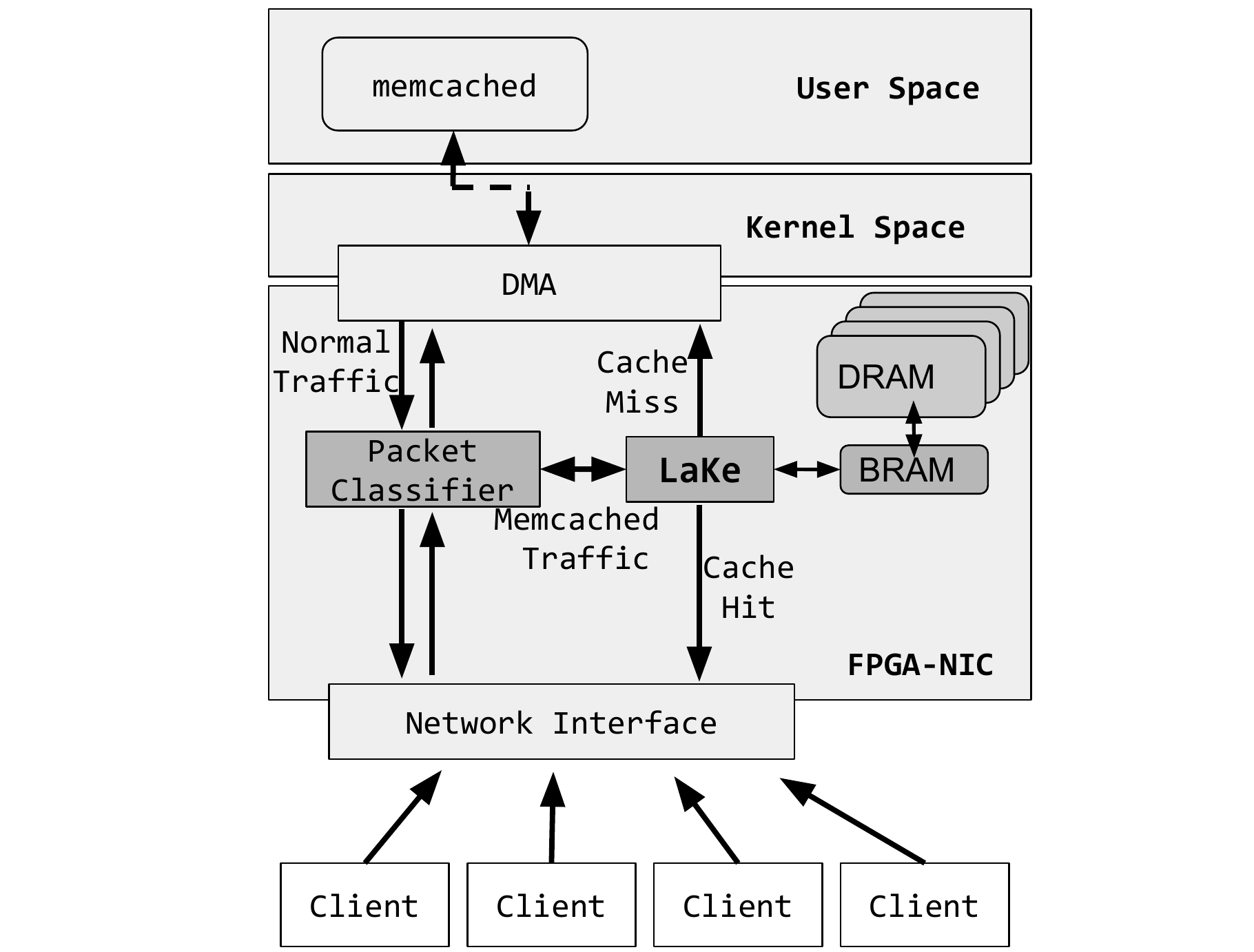}
  \caption{ High level architecture of \acronym. }
  \label{fig:design:hlarch}
  \end{center}
\vspace{-2.5em}
\end{figure}

Single-node memcached servers have been shown in the past to process queries at around 370kqps (query per second) on an Intel Xeon machine~\cite{Lim}, with more modern servers achieving close to a million queries per second . Scaling using consistent hashing algorithms has been shown to improve this throughput by an order of magnitude\cite{Karger_STOC'97}. Large services, such as e-commerce or social networking services, therefore use tens to thousands of data center servers to sustain the query rate they require.

\textbf{\acronym} is a \textbf{La}yered \textbf{Ke}y-value store architecture, focused on memcached. \acronym is FPGA based, and provides significant performance improvement by using multiple layers of cache. Each cache layer provides a trade-off between performance (latency, throughput) and memory size. \acronym runs on a platform that also acts, at the same time, as a NIC or a switch, therefore eliminating the need the cost of adding additional hardware. \acronym provides significant performance improvement compared with host based solutions, reducing by an order of magnitude the number of servers required in the data center. In this section we explore its architecture, as shown in Figure~\ref{fig:design:hlarch}.

\subsection{Background: Key-value Store}
KVS provides simple database functions and API,
such as GET(key) and SET(key, value), issuing a read request and a write 
request, respectively.
KVS stores key-and-value pairs in the host's main memory and storage.
A hash function (e.g., murmur hash, lookup3\cite{Fukuda, Blott13a}, md5, cityhash~\cite{Lim_NSDI'14}) is applied to requested keys, and the hash value is then used to look-up in a hash table the location of the value and meta-data. As a hash table may point multiple keys at the same entry, when an entry is found in the hashed table, the key value in it is compared with the requested key. If both keys are identical, the matched value is returned to the requesting client.

\subsection{High Level Architecture\label{ssec:arch:netfpga}}
The \acronym architecture combines a hardware component and a software component. The software component is the memcached host software, modified to support UDP binary protocol. The hardware component, which is the focus of this paper, is a combined design of a memcached accelerator and a network device running on a single platform.

The architecture of \acronym is shown in Figure~\ref{fig:design:hlarch}. While \acronym can operate either as a switch or a network interface card (NIC), let us assume for clarity that it is used as a NIC. Traffic arrives to \acronym from multiple sources. A packet classifier is used to distinguish between memcached queries and any other type traffic; general traffic will be sent to the host, as in a standard NIC, while memcached queries will be sent to the \acronym module. Queries that are a miss in \acronym's cache and memory, are sent to the host.

\begin{figure}[t]
  \begin{center}
  \includegraphics[width=1\linewidth]{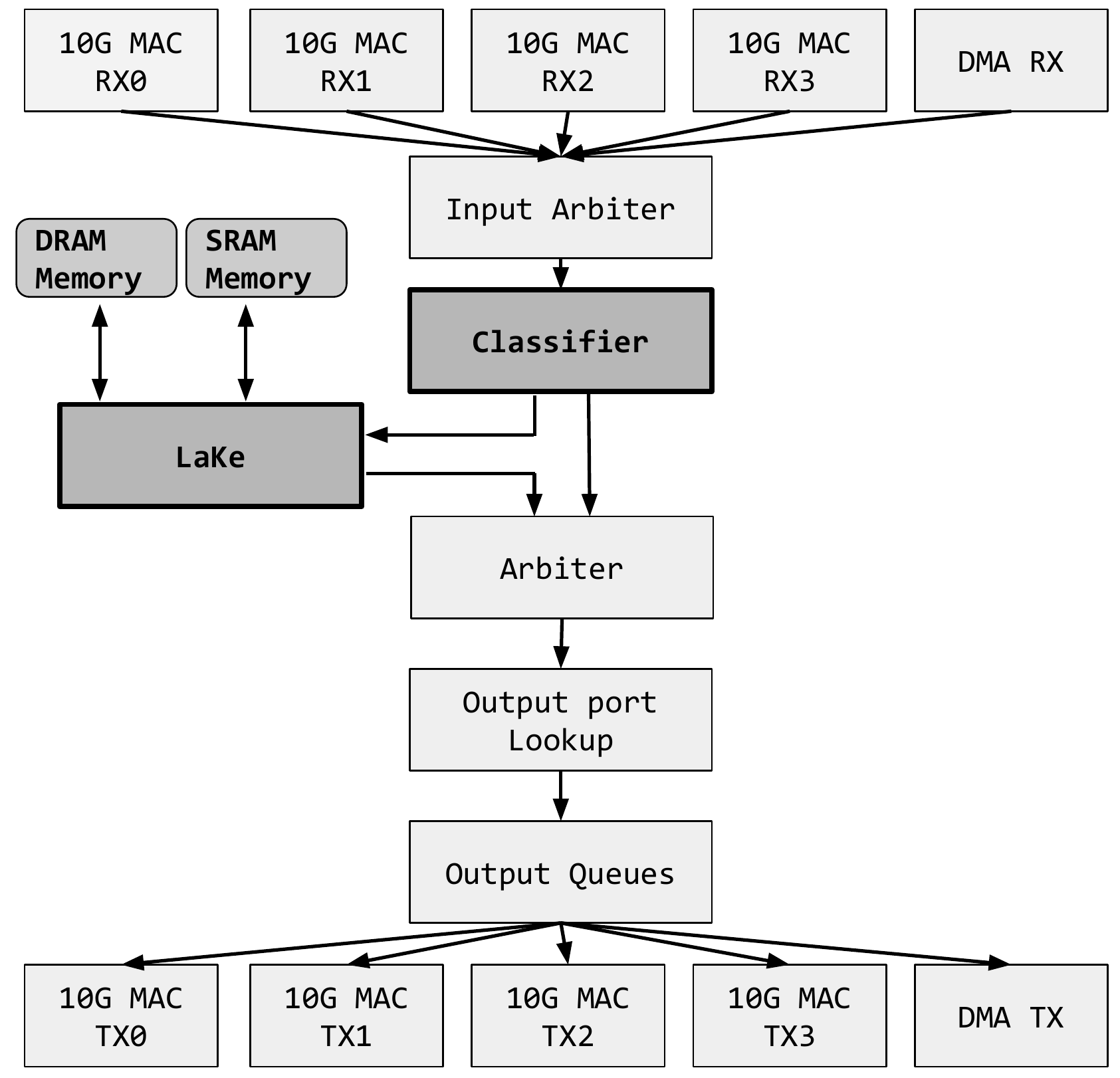}
  \caption{The block design of \acronym integrated 
           with NetFPGA Reference Switch. }
  \label{fig:design:bdiagram}
  \end{center}
\vspace{-2.5em}
\end{figure}

We implement \acronym on the NetFPGA-SUME platform~\cite{netfpga-sume}. The general data plane is based on the NetFPGA Reference Switch project, which can also operate as a NIC, and we amend it with logic enabling memcached operation, as shown in Figure~\ref{fig:design:bdiagram}. Modules unique to \acronym are marked in dark grey. Incoming traffic from multiple ports is fed into the data plane using an arbitration module (Input Arbiter). A packet classifier, unique to our design, identifies the type of the packet, and sends memcached packets to the \acronym module, described later in this section. Non memcached traffic continues in the pipeline, where it is merged (using a second arbiter) with packets returning from the memcached module: both reply packets, going back to clients, and missed queries, forwarded to the host. The destination of the packet is set in an output port lookup module, and packets wait in an Output Queues module to their turn to be transmitted.  
%
%

\subsection{\acronym Module}

\begin{figure}[t]
  \begin{center}
  \includegraphics[width=1\linewidth]{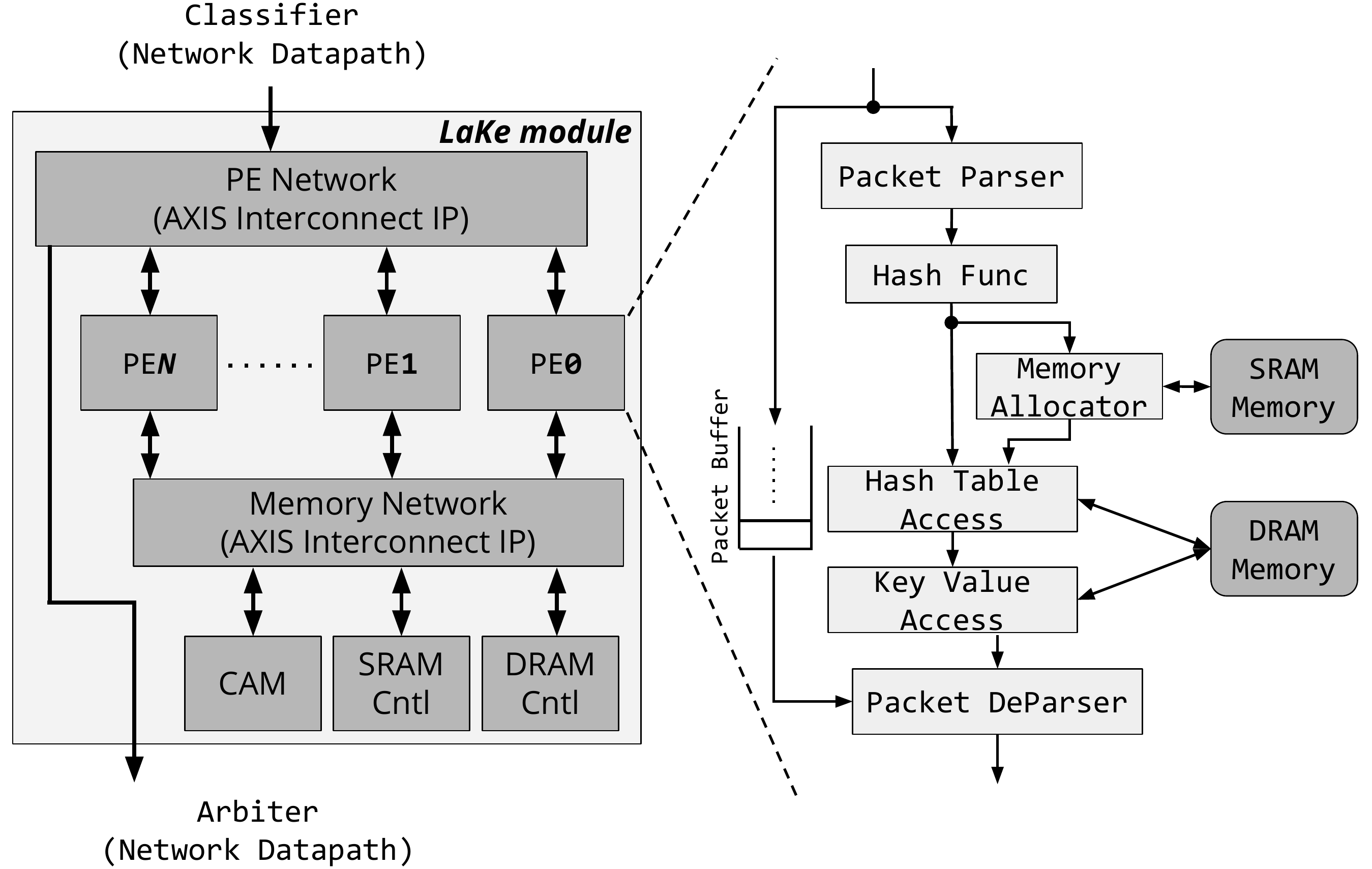}
  \caption{\acronym module architecture. The architecture of each PE is shown on the right.}
  \label{fig:design:pe-arch}
  \end{center}
\vspace{-2.5em}
\end{figure}

To overcome performance bottlenecks and enable scalability across different platforms, \acronym adopts a multi-core processor approach for query processing. The architecture of the \acronym module is shown in Figure~\ref{fig:design:pe-arch}.

Incoming queries are spread between a set of processing elements (PEs), using a PE-network based on an AXIS Interconnect core.
Each PE receives and processes queries. One a query is processed, the PE accesses a shared memory network (using a second AXIS interconnect core). Three types of memories are connected to the memory network: DRAM, containing the hash table bucket and data store chunks (Section~\ref{ssec:dram1}, Section~\ref{ssec:dram2}), SRAM, containing chunk information (Section~\ref{ssec:sram}), and CAM, serving as a look up table (LUT) for retrieving key-value pairs (Section~\ref{ssec:lut}). 

As a new query arrives, the PE parses the packet and extracts the command, key and value. Next, the hash of the extracted key is calculated. In our implementation, CRC32 is used as the hash function. The hash value serves as a pointer to an address in the DRAM, holding a descriptor (hash table bucket) pointing to the key-value pair in the memory.

Upon a SET command, both the hash table and the key-value pair data are updated in the DRAM. 
If a SET command arrives with a new key, the PE assigns it to a chunk using a list of free descriptors stored in the SRAM and pointing to empty chunks. 

If a key exists in \acronym's memory, it is considered a \textit{hit}, and a reply is prepared in the \textit{Packet Deparser} and returned to the client. Otherwise, the request is be forwarded to the host machine through the switch datapath and using a DMA engine~\cite{netfpga-sume}.

\subsection{Hash Table}\label{ssec:dram1}

The hash table is used to store descriptors pointing from a hashed key
to the address in memory of the actual key-value pair. As such, it is a critical component in the design. 
The data structure of the descriptors in the hash table is shown in Figure~\ref{fig:design:htstrct}.
The descriptor size is 64bit, which is performance optimized: the DDR3 SoDIMM on the board uses a bus width of 64bit and a burst size of eight, which leads in turn to a bus width from the DDR3 controller of 512bit. This allows in a single access to read eight descriptor entries, enabling 8-associativity. To reduce the number of accesses to the DRAM, a key's length is compared to the key's length in the descriptor, and only if they match the PE attempts to access the DRAM and read the key-value chunk.

\begin{figure}[t]
  \begin{center}
  \includegraphics[width=0.8\linewidth]{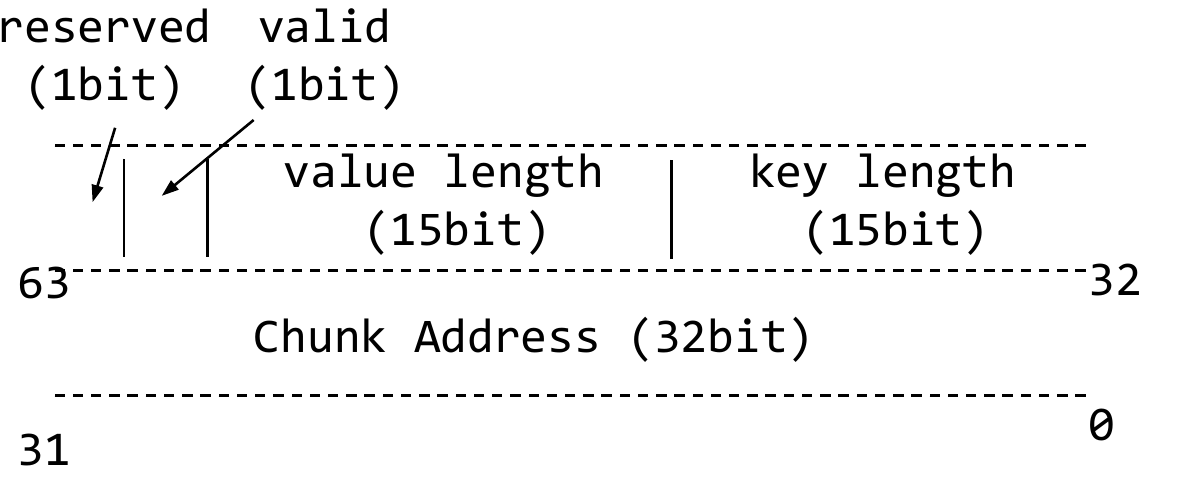}
  \caption{ Hash table format. Fixed size entries are used. }
  \label{fig:design:htstrct}
  \end{center}
\vspace{-1.5em}
\end{figure}

\begin{figure}[t]
  \begin{center}
  \includegraphics[width=1\linewidth]{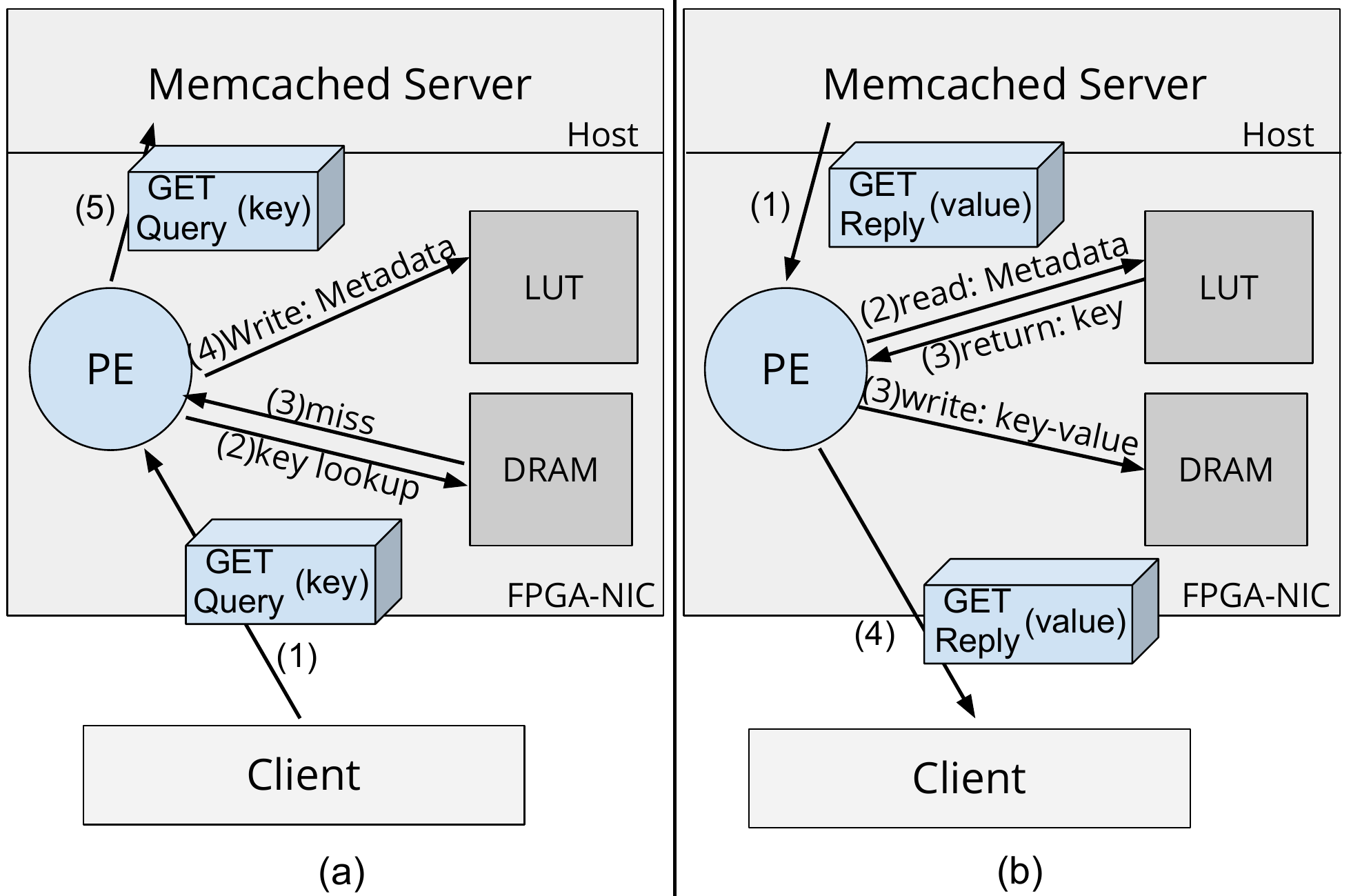}
  \caption{The request-response process of a query miss in \acronym. (a) When a GET request is a miss in the cache and the DRAM of \acronym, the requested key is learned in a LUT, associated with source UDP port number and memcached opaque field, and the query is forwarded to the host. (b) As a GET reply is returned from the server, the key is retrieved from the LUT and the value paired with the key is stored in the DRAM. The GET reply is sent to the client.}
  \label{fig:design:htstrct}
  \end{center}
\vspace{-1.5em}
\end{figure}

\subsection{Memory Management}\label{ssec:dram2}\label{ssec:sram}

Memcached builds upon a slab allocator to efficiently use the memory~\cite{Rajesh_NSDI13, memcached}. This approach is also taken in hardware based designs, as well as in \acronym, enabling to handle variable key- and value-length.

A slab allocator is implemented using an SRAM-based memory, storing addresses of unused chunks. To reduce access time to the SRAM, \acronym uses a small cache (implemented as a FIFO), which pre-loads the next available addresses from the memory. 
The number of entries in the SRAM can be calculated using the following formula: 
$ \sum_{k=i}^{n} S_{k} N_{k} \leq C_{mem} $, where
$S_{k}$, $N_{k}$ and $C_{mem}$ denotes the size of chunk, the number of chunks and SRAM capacity, 
respectively.
We use multiplications of 64B as slab size, and support 64B, 128B, 256B and 512B chunk sizes in our prototype. The minimum size slab is determined by the width of the memory network datapath: 512bit.

{\bf DRAM access: }
Random access to the DRAM has a non negligible and variable latency, which can stall PEs. 
To attend to this latency we integrate a small cache (e.g., 64B cache line, write through, direct-map, total capacity 64kB). 
Without the cache, we measure the DRAM controller's access latency (using Xilinx MIG, running at 933.33MHz) to be around 115ns in a zero load test, and to be up to 650ns under high load. 

{\bf PE scalability: } 
\acronym applies a modular, scalable approach to KVS acceleration.
The number of PEs supported by the design starts at one and scales up, with five PEs sufficient to support per-port full line rate.
Beyond physically implementing a variable number of PEs, \acronym also allows to control on-the-fly the number of PEs used, balancing workload and power efficiency.


\subsection{Memcache Protocol\label{ssec:lut}}

Memcached systems~\cite{Rajesh_NSDI13} generally use the memcache protocol.
There are two memcache protocol variants: ASCII based and binary based. Our implementation uses the binary variant.
The challenge using the memcache protocol is that key-value pairs cannot be identified in responses from the host.
For instance, a GET request missed in the hardware and sent to the host will have a query response returning with the value but without the key. Thus, cache systems cannot handle only response packets; it is required to learn and save request query's information. 

To associate a key with a returning value, we use memcache protocol's opaque field and source UDP port number. Memcache protocol uses a 32-bit opaque field, and memcached systems use the same opaque value in both request and reply.
We use a lookup module to match returned values from a host with their paired keys.
The LUT is implemented using a CAM, where we query using the opaque value and the source UDP port, and the reply is the original query's key. 
The keys are updated every time a GET query is a miss in the hardware and forwarded to the host.

\section{Implementation\label{sec:imp}}
Our target board is NetFPGA-SUME\cite{netfpga-sume}, which is equipped with Xilinx Virtex-7 690T FPGA, 8GB DDR3 SDRAM modules ($4$GB$\times 2$, upgradable to $16$GB$\times 2$), three QDRII SDRAM modules (27MB) and more.
The NetFPGA Reference Switch project is the baseline datapath, integrated with the memcached subsystem, as shown in Figure~\ref{fig:design:bdiagram}.
The project is implemented in Verilog HDL, using Xilinx Vivado 2016.4 design flow. 

The current implementation of \acronym supports up to thirteen PEs, though only five PEs are required to achieve full line rate. The limitation on the number of PEs is due to the number of slave interfaces available on the AXI-Steam interconnect cores, used by the PE interconnect and the memory interconnect  (16 slaves, where SRAM, DRAM and CAM must always be connected). The core clock frequency is 200MHz.

Using five PEs, the fully implemented prototype consumes only 35.65\% of the Block RAM (BRAM) and 52.33\% slice utilization\footnote{Note that the NetFPGA Reference Switch uses 13.91\% of the Block RAM (BRAM) and 17.9\% slice utilization}. On a higher end FPGA this number is significantly smaller, allowing scalability to higher data rates.

\subsection{Integration with NetFPGA Datapath}
\acronym is integrated with the NetFPGA switch/NIC datapath as shown in 
Figure~\ref{fig:design:bdiagram}. \acronym gives priority to normal traffic over memcached traffic; if the memcached packet rate is high and over-subscribes the cache sub-system, instead of throttling normal traffic \acronym drops memcached packets. Consequently, normal traffic is not affected by \acronym's performance.   
Outgoing packets from \acronym module go through an arbiter, which
arbitrates between memcached packets and normal packets, forwarding them to an \textit{Output Port Lookup} module.

\acronym is a caching system of memcached, meaning that SET and DELETE requests need to be updated in the host's memory. Consequently, SET and DELETE requests are copied to both paths within the FPGA: to \acronym and to normal traffic path.
Specifically, while a SET request to \acronym updates or adds new cache contents to the shared-cache and DRAM module on NetFPGA, the SET request sent through the normal traffic path updates or adds new content to the host memory, running the memcached server (software).
In addition, a reply from the host to a GET request missed in the hardware, also goes through the normal traffic path and to \acronym's module, updating the local cache contents.

\subsection{Scalability}

\acronym scales up both in throughput and resources.

\begin{figure*}[t!]
\hspace{-3mm}
	\begin{minipage}[t] {.331\textwidth}
	\centering
		\includegraphics[height=45mm]{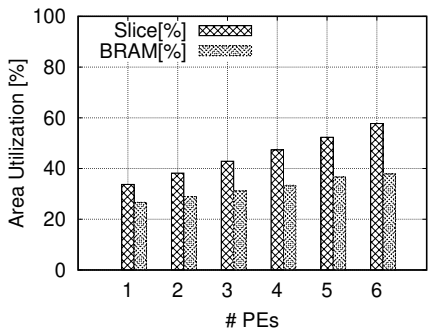}
		\caption{ The area utilization of \acronym implemented on NetFPGA SUME. }
		\label{fig:eval:area}
	\end{minipage}
	\begin{minipage}[t] {.331\textwidth}
	\centering
		\includegraphics[height=45mm]{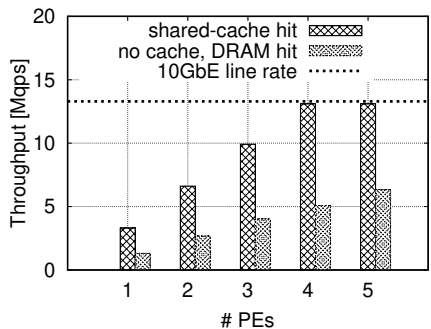}
		\caption{ Throughput as a function of number of PEs. }
		\label{fig:eval:tput}
	\end{minipage}
	\begin{minipage}[t] {.331\textwidth}
	\centering
		\includegraphics[height=45mm]{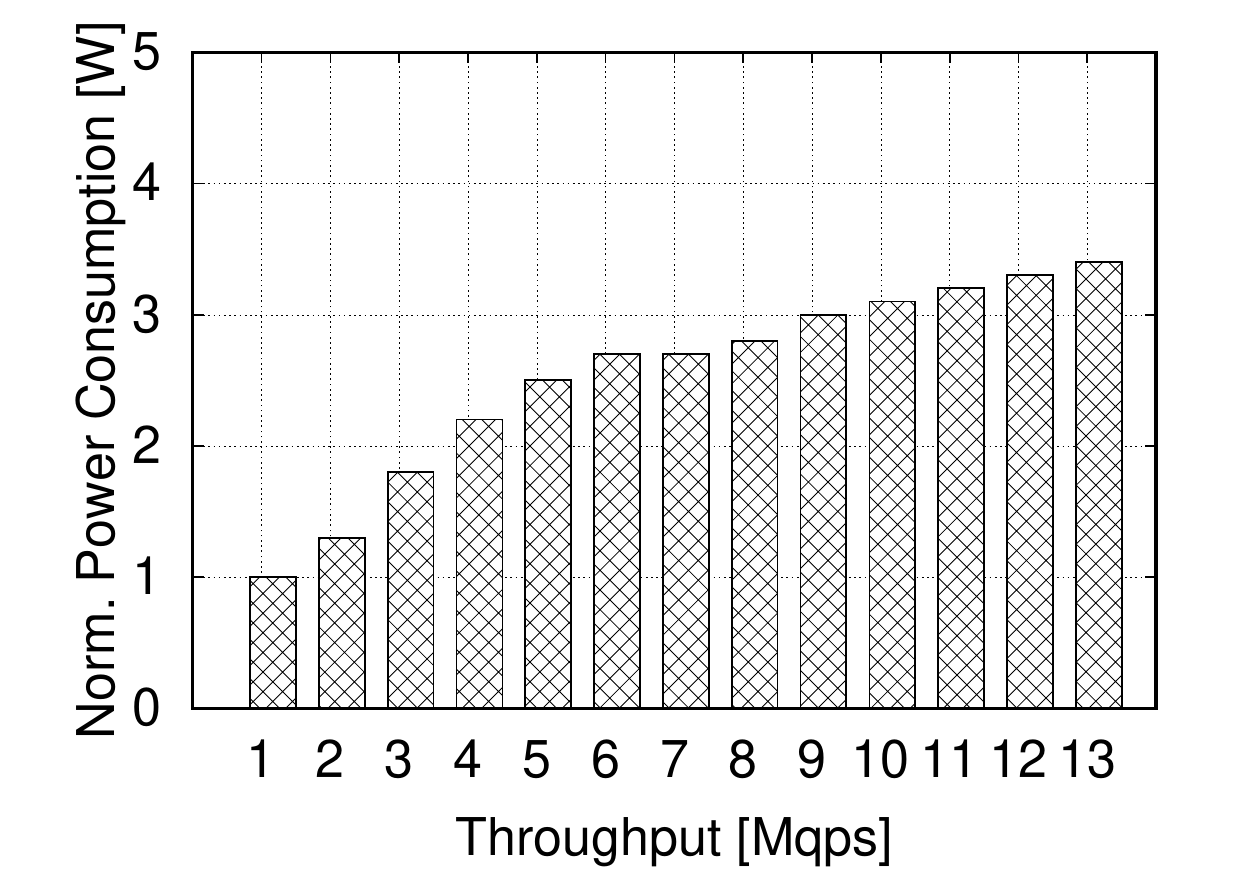}
		\caption{ Throughput vs power consumption. }
		\label{fig:eval:performance}
	\end{minipage}
	\vspace{-8mm}
\end{figure*}

{\bf Area and Resources:}
We implemented up to six PEs while maintaining 200MHz core frequency, as shown 
in Figure~\ref{fig:eval:area}.
Each PE utilizes around 3\% of chip slices and 2\% BRAMs. These values include also the interconnection networks, as  
each PE is connected with both PE interconnect and memory switch.  The small overhead in resources taken by each PE enables scaling the number of PEs used by \acronym with little effect on resource consumption. 

{\bf Throughput:}
We evaluate the throughput scalability of \acronym using OSNT\cite{Antichi_Network'14}.
First, the cache is warmed using a SET request. Next, OSNT generates GET requests, matching the warmed cache, using a 4B key, and returning an 8B value. The throughput scalability as a function of the number of PEs is shown in Figure~\ref{fig:eval:tput}. As the figure shows, 
\acronym can handle up to $13.1$Mqps using five PEs, when the queries are \textit{hit} in the shared-cache in front of the  DRAM. Each PE processes up to $3.3$Mqps.
The bottleneck on throughput growth is the interconnect core and memory bandwidth. The throughput grows linearly with the number of PEs until reaching these bottlenecks. On a platform with more memory interfaces, or with a higher speed memory, a higher throughput can be reached.

%
%

\section{Evaluation}\label{sec:eval}
The evaluation of \acronym covers two aspects: absolute performance, 
and the exploration of design trade-offs. The evaluation results are summarized in Table~\ref{tab:eval}.

\subsection{Absolute Performance}

\acronym is not only feasible, but also achieves high performance.
We evaluate the absolute performance of \acronym based on several performance metrics: throughput, latency and power efficiency. We compare the performance with memcached (v1.5.1), a software implementation, and Emu's memcached implementation~\cite{Nik_ATC'17}, a hardware based memcached system using the binary protocol. Emu is selected as it is comparable, being as open-source available on NetFPGA-SUME.

%
%

%
%
%
%

\subsubsection{Test Setup}
The server uses Intel Core i7-4770 CPU, 64GB RAM, running Ubuntu 14.04 LTS (Linux kernel 3.19.0) and NetFPGA-SUME card.
OSNT~\cite{Antichi_Network'14} is used for traffic injection. A 10GbE port is connected to \acronym-side card.
GET requests including 4B key and 8B value are injected at 10Gbps. Throughput is measured on a second granularity.
For comparison with software-based memcached, we amended the memcached software to support 
binary protocol over UDP. 


\subsubsection{Maximum Throughput}

\acronym achieves a throughput of $13.1$Mqps (query per second) when all the queries are \textit{hit} in the shared-cache, as shown in Table~\ref{tab:eval}. 
This is $\times 6.7$ improvement compared with Emu~\cite{Nik_ATC'17}.
While Emu does not use an external memory module,
\acronym can utilize the large DRAM capacity to handle cache misses. 
Future work will evaluate maximum throughput using more than six PEs.



%

\subsubsection{Latency}

We use an Endace DAG card 10X2-S to measure queries' latency. 
A software-based client is used to generate queries, and the DAG measures the isolated latency of \acronym, client excluded.
Despite using DRAM, \acronym's latency is almost equivalent to Emu, thanks to the small shared-cache (64kB) in front of the DRAM.
When queries are \textit{hit} in the cache, the latency is about 1.1$\mu$s per query. When queries are \textit{miss} in the shared-cache, and \textit{hit} in the DRAM, the latency is 5.6$\mu$s.

\subsubsection{Power Efficiency}

We use a wall power meter to measure power consumption.
We calculate power efficiency as 
$ E = W/T $, where $E$, $W$ and $T$ denote power efficiency,
power consumption and throughput, respectively.
\acronym achieves 242.962 kqps/Watt using five PEs at full line rate. This is
$\times 5.1$ improvement compared with Emu.

We investigate dynamic power consumption by measuring how power consumption varies as a function of throughput.
Power consumption is normalized to $0$W under zero load, i.e. the static power consumption. 
The dynamic power consumption takes up to 3.4W on our LaKe modules (as shown in Figure~\ref{fig:eval:performance}), while software based memcached consumes a maximum of 58.2W dynamically. Thus, \acronym improves dynamic power consumption drastically.

%
%
\begin{table*}[t!]
\centering
\caption{Performance comparison.}
\begin{tabular}{c|r|r|r}
\hline\hline
System  & Average latency [$\mu$s] & Throughput [Mqps]& Power efficiency [kqps/Watt.]\\
\hline
memcached(software) & 238.84 & 0.962 & 9.938 \\
Emu (hardware)\cite{Nik_ATC'17} & 1.21 & 1.932 & 47.121 \\
\acronym (shared-cache) & 1.16 & 13.120 & 242.962 \\
\hline
\end{tabular}
\label{tab:eval}
\vspace{-2.1em}
\end{table*}

\subsection{Design Trade-Offs}\label{sec:eval-design}

In the previous section we have seen that there is no silver bullet to power efficiency: it depends on the load. In this section we focus on design trade-offs that affect the power efficiency of \acronym.

So far the evaluation used a fully-featured \acronym: using BRAM, SRAM and DRAM. Here, we check the contribution of each.
Note that for following trade-offs discussion we employ a single DRAM module (4GB) which utilizes both hash table region (2GB: 268M entries) and a data store region (2GB: 33M entries as 64B chunk). When we use BRAM instead of DRAM, the number of hash table entries and data store entries are 4096 entries and 512 entries, respectively. We also employ two SRAM modules (total 18MB) to manage free-list on slab allocation. When we use BRAM instead of SRAM, the number of free-list addresses stored is 144 entries. 

When only the BRAM is used, and the SRAM and DRAM memory controllers are taken out, the maximum power consumption of \acronym is 16W including NetFPGA-SUME card, and the maximum throughput is $13.1$Mqps. Under these circumstances we use 1k entry BRAM-based cache as hash table and data store instead of a DRAM, and use BRAM-based FIFO as slab allocator instead of an SRAM.

Adding the SRAM adds 6W and holds 4.7M chunk addresses, which are updated when a DELETE operation moves a specific chunk to the free list. A BRAM-based FIFO placed in front of the SRAM is used to hide SRAM access latency, but is shallow in comparison with the SRAM.
One can therefore trade the 6W SRAM power consumption with the number of available chunks on \acronym.

Adding the DRAM, but not the SRAM, adds 4W and around 33M data store entries and 268M hash table entries, using only a data-store chunk size of 64B.
DRAM access latency directly affects its throughput. As shown in Figure~\ref{fig:eval:tput}, the maximum throughput is 6.3Mqps when five PEs are implemented, lower than using the shared-cache. More PEs or using a BRAM based cache are required to achieve line-rate.
The trade off using DRAM is not only power consumption to number of hash table and data store entires, but also throughput.

The fully featured design has 1.1$\mu$s latency on a hit, and achieves $13.1$Mqps 10GbE line rate processing, but has a cost of additional 10W to the power consumption for external memories. 
This means that one needs to consider power consumption trade off with performance and memory size.
%
%
%

\section{Related Work}\label{sec:rwork}

Hardware-based KVS has been actively researched along with the rise of cloud computing~\cite{Chalamarasetti, Labasani, Jae, Blott13a, Blott15, Tokusashi_HOTI'16, Tokusashi_MICRO'17, Lim, John, Fukuda, Xu_VLDB'16}.
It was previously shown that offloading KVS into dedicated hardware, such as 
FPGA or ASIC, benefits in terms of latency, throughput and power efficiency.
Although hardware based memcached appliances~\cite{Blott13a,Blott15} 
were shown to achieve 10GbE throughput, 
the cache capacity was small, limited by physical resources and FPGA IO constraints.
In contrast, our work focuses on a layered cache architecture, benefiting from an integration with the host machine. As long as queries are a hit in the FPGA-NIC, the CPU load is significantly reduced.

KV-Direct~\cite{Li_SOSP'17} demonstrated a switch achieving a high query rate (e.g., $\sim$180Mqps), but was limited to a proprietary solution, using 8B query size, batching multiple queries in a single packet, and processing vector queries. \acronym supports the highly popular memcahce protocol and different slab sizes, offering a far richer feature set.

Programmable switch ASICs (e.g., Barefoot Networks Tofino), are providing extreme high performance. KVS on programmable switches was demonstrated using cache capacity resources in~\cite{Jin_SOSP'17}. Until such devices have significant memory resources attached, they will not be able to completely offload KVS applications.

Energy efficiency in KVS was also researched in software-based solutions ~\cite{Lim_NSDI'14, Li_ISCA'15, Li_TOCS'16, Li_MICRO'16, Lim_SIGMOD'17, Liu_ASPLOS'17, Shim_JSS'17}.
Software-based solutions improve performance and power efficiency by using the CPU more 
effectively. 
However, software-based solutions do not improve latency and power consumption, 
while \acronym has improved latency, throughput and energy efficiency dramatically.

\section{Conclusion}\label{sec:conc}

While network bandwidth is ever increasing, computing performance is leveling off.
Performance requirements, however, need to be balanced with power efficiency.
In this paper, we presented \acronym, a new architecture for
energy efficient KVS.
\acronym presents a multi-core, multi-level cache architecture, that balances throughput, latency and power efficiency: achieving $\times 5.1$ better energy efficiency than existing FPGA-based memcached systems and $\times 6.7$ better throughput, maintaining almost the same latency. \acronym does all that without giving up memcached functionality. 
\acronym demonstrates the cost of performance: balancing throughput and memory size against power consumption. 



\bibliographystyle{IEEEtran}
\bibliography{fpl2018}


\end{document}